# Topological transport in Dirac electronic systems: A concise review

Hua-Ding Song [1], Dian Sheng [1], An-Qi Wang [2], Jin-Guang Li [1], Da-Peng Yu [1,3,4], and Zhi-Min Liao [1,3,†]

[1] State Key Laboratory for Mesoscopic Physics, School of Physics, Peking University, Beijing 100871, China
[2] Academy for Advanced Interdisciplinary Studies, Peking University, Beijing 100871, China
[3] Collaborative Innovation Center of Quantum Matter, Beijing 100871, China
[4] Department of Physics, South University of Science and Technology of China, Shenzhen 518055, China

Various novel physical properties have emerged in Dirac electronic systems, especially the topological characters protected by symmetry. Current studies on these systems have been greatly promoted by the intuitive concepts of Berry phase and Berry curvature, which provide precise definitions of the topological orders. In this topical review, transport properties of topological insulator ($Bi_2Se_3$), topological Dirac semimetal ($Cd_3As_2$) and topological insulator-graphene heterojunction are presented and discussed. Perspectives about transport properties of two-dimensional topological nontrivial systems, including topological edge transport, topological valley transport and topological Weyl semimetals, are provided.

## 1. Introduction

Inspired by the notion of "topological order", the exploration for novel classifications of matter phases becomes an intriguing theme in condensed-matter physics. Over three decades, the discovery of quantum Hall effect (QHE) enlightened investigations on topological phases of matters[1-6]. Assisted by the powerful concepts of Berry phase and Berry curvature, precise definitions of topological orders are achieved. Berry phase is the phase generation during an adiabatic evolution process with steady external parameters producing a loop in the parameter space[1,2]. The Berry phase can be written as the integral of the Berry curvature[2]. Moreover, the integrals of the Berry curvature over closed surfaces are always employed to identify topological nontrivial phases, known as quantized Chern numbers[2]. Hence, theoretical predictions for nontrivial matters emerged endlessly for more than a decade, and novel quantum systems were continuously revealed. Such topological nontrivial systems exhibit fascinating properties, especially in transport experiments.

The QHE was discovered by Klaus von Klitzing *et al*, where the Hall conductivity of a two-dimensional (2D) system was found to be exactly quantized under a strong magnetic field[7-9]. Thouless *et al.* firstly employed topological considerations to explain the QHE[5]. From their milestone work, researchers began to realize that the electronic topological structure of quantum materials may be directly related to the transport behaviors. Various materials with novel topological order have already been revealed. Graphene, the first 2D material containing Dirac fermions, has been proved to possess a Berry phase of π, leading to half-integer QHE observed in transport measurements[10-16]. Topological insulators (TIs) like $Bi_2Se_3$ and $Bi_2Te_3$, where strong spin-orbit coupling creates topological insulating electronic phases and non-zero integer Chern number, are very potential platforms for spintronics and next-generation information devices[17-33]. Gapless surface states protected by time-reversal symmetry exist on the surface of these materials, and less-dissipative transport can be achieved by the surface states, leading to information highway in the future. Similar to graphene system, the surface states of TIs also behave as 2D Dirac fermions with π Berry phase[18-21]. Thus, Hall conductivity of the surface states is half-integer quantized as well. However, the parallel connection of the top and bottom surfaces may necessarily carry an integer QHE in real measurements[23]. Rui Yu *et al.* proposed that quantum anomalous Hall effect (QAHE) can be realized in magnetized TI system[25], which was confirmed by Cui-Zu Chang *et al*[26,27]. Recently, topological semimetals became another focus of condensed-matter physics. The topological Dirac semimetals (for example, $Na_3Bi$, $Cd_3As_2$) are natural three-dimensional counterpart of graphene, with bulk linear-dispersion Dirac fermions and Fermi arcs surface state [34-39]. The unique chiral character of the Dirac fermions in topological Dirac semimetals generates abundant physical entities, such as chiral anomaly induced charge pumping behavior[39]. Novel transport properties, like anomalous negative magnetoresistance and surface state assisted Aharonov-Bohm oscillations[40,41], were successively discovered.

[†] Corresponding author. E-mail: liaozm@pku.edu.cn

In this topical review, we generally introduce the experimental investigations about transport behaviors in topological nontrivial systems, mainly based on our own works. We have reported abundant phenomena originating from surface states in topological insulators, topological Dirac semimetal and monolayer graphene. In section 2, we demonstrate the enhanced photothermoelectric effect by TI helical surface states[42], and the quantum oscillations from TI nanoribbon sidewalls[43]. In section 3, our investigations on the topological Dirac semimetal $Cd_3As_2$, including negative magnetoresistance induced by chiral anomaly[40], Aharonov-Bohm oscillations in $Cd_3As_2$ nanowires and two-carrier transport in $Cd_3As_2$ nanoplates are summarized[41,44]. In section 4, we exhibit the novel tunneling behavior in the vertical heterojunction formed by graphene and TI[45]. In section 5, we will review other significant theoretical and experimental progresses on the topological edge state transport, the valley transport and the topological Weyl semimetals. Finally, we give perspectives and conclusions in section 6.

## 2. Transport properties of topological insulator $Bi_2Se_3$

### 2.1 Enhanced photothermoelectric effect of 2D topological surface states

Here we review the photothermoelectric effect in topological insulator $Bi_2Se_3$ nanoribbons. $Bi_2Se_3$, as a strong three-dimensional topological insulator, has a bulk energy gap of ~0.3eV and helically conducting chiral surface states[22,24]. The 2D surface states could be excited to be spin-polarized, by means of circularly polarized light. Considering spin-momentum locking in topological insulators, the spin-polarized surface states have oriented motions, which could be accelerated by the temperature gradient, resulting in enhanced photothermoelectric effect[46,47].

The $Bi_2Se_3$ nanoribbons were synthesized via chemical vapor deposition (CVD) method. To better characterize the crystalline quality and study layered microstructure of the nanoribbons, the as-synthesized $Bi_2Se_3$ nanoribbons were prepared to be cross-sectional on a $SrTiO_3$ substrate (Fig. 1a). Measured from the cross section along the [100] zone axis, the diffraction pattern indicates that the nanoribbon is single crystalline (Fig. 1b) with the basic unit of quintuple layer of -Se-Bi-Se-Bi-Se- sequentially (Fig. 1c). The ARPES spectrum acquired at room temperature confirms the existence of TI surface states (Fig. 1d).

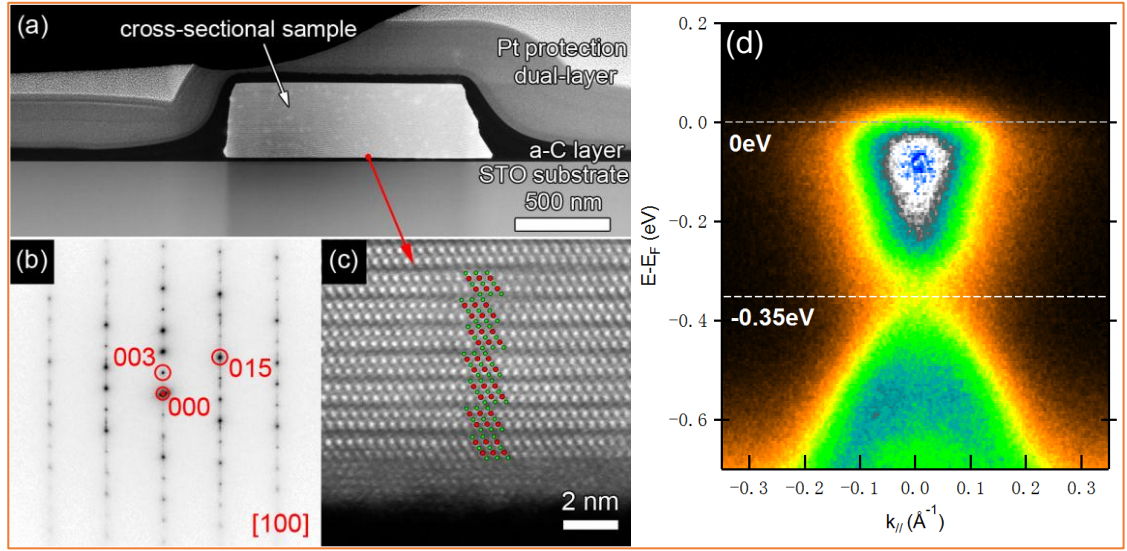

**Figure 1.** (**a**) Cross-sectional HAADF-STEM image of a $Bi_2Se_3$ nanoribbon. (**b**) Diffraction pattern acquired from the cross-section along the [100] zone axis. (**c**) High resolution HAADF-STEM image of the nanoribbon surface, where the quintuple layer layer is indicated. (Bi represented by red dots; Se represented by green dots.) (**d**) Room-temperature ARPES spectrum for a $Bi_2Se_3$ nanoribbon.[42]

Individual $Bi_2Se_3$ nanoribbons were fabricated to contact with Cr/Au electrodes on a Si substrate with 300nm $SiO_2$ layer via electron beam lithography techniques, as shown in Fig. 2a. To study the surface state enhanced photothermoelectric effect in $Bi_2Se_3$ nanoribbons, the voltage response to circularly polarized light illumination were measured. Right-circular polarization (RCP), linear polarization (LP), and left-circular polarization (LCP) of the incident laser were achieved *via* tuning the λ/4 waveplate. The incident laser was along y-z plane, where $\theta$, the relative orientation of incident light and z axis, was chosen to be 0° and 30° (Fig. 2b). Considering that most notable voltage response was detected when illuminating near the electrodes, the incident laser in our experiment was illuminated near the measurement electrodes. As $\theta = 0°$, the voltage vs time curves of RCP, LP, LCP show a little difference, as a result of that the vertically incident laser cannot induce in-plane surface states to be spin polarized, thus no additional surface state related photocurrent can be excited.

As $\theta = 30°$, it is found that the RCP and LCP light have different effect on photothermoelectric output. When the laser spot is near the electrode connected to the voltmeter positive terminal (the positive electrode), the RCP light induces an obviously enhanced voltage signal, the voltages induced by LCP and LP are similar (Fig. 2c). However, when the laser is illuminated near the voltmeter ground terminal (the negative electrode), LCP induced photothermoelectric effect is notable, while voltage response of RCP is almost the same as LP (Fig.2d).

The phenomena above are dominated by the light polarization selective rule according to the helical character of the carriers (Fig.2e). We define the electrons in the Dirac cone with positive slope as spin-up, while those in the branch with negative slope correspond to spin-down states. The quantum numbers of the orbital angular momentum of each branch are +1 and -1, respectively[48-51]. A photon can selectively excite the surface states to the conduction band states ($l=0$) by transferring angular momentum[48,51]. If spin-polarized surface electrons have the same direction as the electrons driven by temperature gradient, the generated voltage would be enhanced. Conversely, if the additional surface electrons have the opposite motion direction with the electrons driven by temperature gradient, there would be little impact on the total generated voltage.

When the laser illumination on $Bi_2Se_3$ is near the positive electrode, the $\sigma-$ photons (RCP) only excite the surface electrons with $m_l = 1$ to conduction bands ($l = 0$), while the hopping of $m_l = -1$ electrons is forbidden, resulting in additional spin-up surface states. Considering the spin-momentum locking in $Bi_2Se_3$, the spin-up surface electrons have a translational direction along as +x axis[52], which is parallel to the direction of temperature gradient. Naturally, additional surface electrons would be accelerated and enhance photothermoelectric effect. Contrarily, the LCP would hardly enhance photovoltage as a result of anti-parallel relationship between the momentum of spin-down surface electrons and temperature gradient. A similar analysis can be made for the laser illumination position near the ground electrode, in this case, the LCP would induce significant enhancement of photothermoelectric effect.

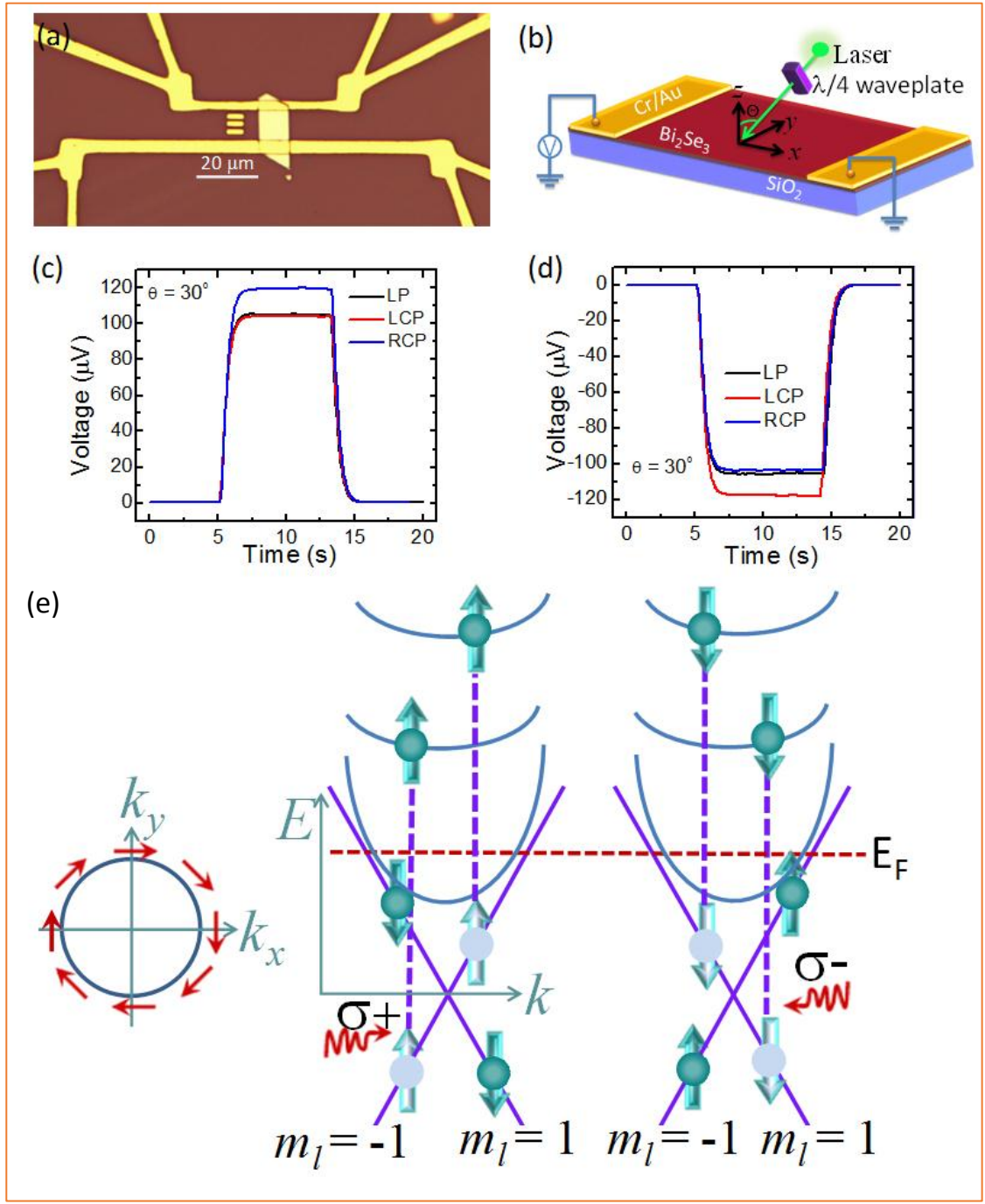

**Figure 2.** (**a**) Optical image of a typical $Bi_2Se_3$ nanoribbon PTE device. (**b**) Sketch of the measurement set up. (**c**) The photo-voltage response measured when the laser illumination on $Bi_2Se_3$ near the positive electrode. (**d**) The photo-voltage response measured when the laser illumination on $Bi_2Se_3$ near the ground electrode. (**e**)The physical mechanism for the light polarization-selective transition. [42]

**2.2 Side-wall surface states associated quantum oscillations**

Three-dimensional topological insulators have been demonstrated to possess rich physical connotations[17-24]. $Bi_2Se_3$ is always considered as a perfect platform to explore the novel properties of 2D Dirac fermions on its surface. However, even containing a large bulk energy gap, the conductance contributed from bulk is still notable due to Se vacancies in $Bi_2Se_3$, making identification of the surface states by transport measurements a great challenge. In this section we review our investigations on prominent SdH oscillations originating from the surface Dirac fermions in the sidewalls of $Bi_2Se_3$ nanoplates. Importantly, the SdH oscillations from the sidewalls appear with a dramatically weakened magnetoresistance background, offering a direct path to detect the 2D surface states when the coexistence of bulk and surface conduction is inevitable.

The $Bi_2Se_3$ nanoplates were prepared by the vapor-liquid-solid mechanism, using gold as catalyst. Electrodes of Cr/Au (10 nm/130 nm) were deposited by electron beam evaporation for transport measurements (Fig.3b). When applying an in-plane magnetic field, obvious SdH oscillations can still be observed on a smoothed magnetoresistance background (Fig.3c), the oscillation features are gradually smeared out by increasing temperature (over 20K). Such in-plane field induced SdH oscillations were rarely reported before, and apparently, this phenomenon does not originate from top/bottom surfaces. These results indicate that the most possible origination is the surface states on two sidewalls.

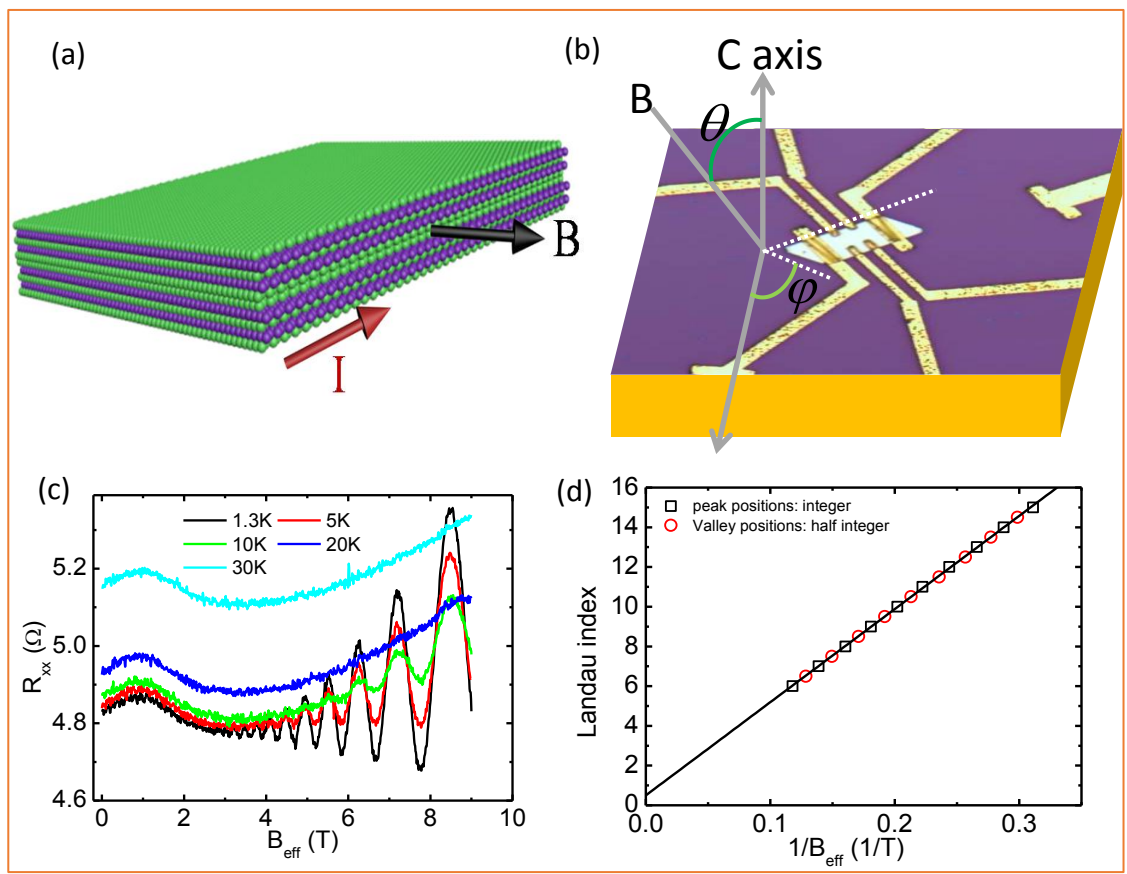

**Figure 3.** (**a**) Schematic of the in-plane magnetic field referring to the sidewall (Se: green spheres; Bi: purple spheres).(**b**) Optical image of a typical $Bi_2Se_3$ nanoplate device and a schematic of the orientation of the applied magnetic field. The length between the two inner longitudinal voltage leads, and the width of the two Hall

voltage leads of the Bi$_2$Se$_3$ nanoplate device are 1.48 μm and 2.26 μm, respectively. (**c**) Plot of longitudinal resistance $R_{xx}$ *versus* $B_{eff}$ under in-plane magnetic field at different temperatures. (**d**) Landau-level fan diagram for the Shubnikov-de Haas oscillation in R$_{xx}$ at the lowest temperature 1.3 K.[43]

Considering the Onsager semiclassical quantization relation, a linear relationship can be acquired: $N_{min} = \mathcal{K}\frac{1}{B} - (\gamma - \frac{1}{2})$, where $N_{min}$ indexes the $N_{th}$ minimum in conductance, $\mathcal{K}$ is a constant, $\gamma - \frac{1}{2}$ is related to the Berry phase by $\gamma - \frac{1}{2} = -\frac{\beta}{2\pi}$. Based on such relation, the Landau level fan diagram can be established (Fig.3d). The extrapolated Landau index at extreme field limit (1/B~0) is about 0.5, acquired from Landau level fan diagram, indicating an additional Berry phase π in the system. This is a direct evidence of novel 2D Dirac fermions. We substantially demonstrated that the oscillation behavior derived from Landau level formation in 2D Dirac fermions. Since applied in-plane field cannot induce such formation in top/bottom surface, the sidewalls are the only possible origin for the quantum oscillation we observed.

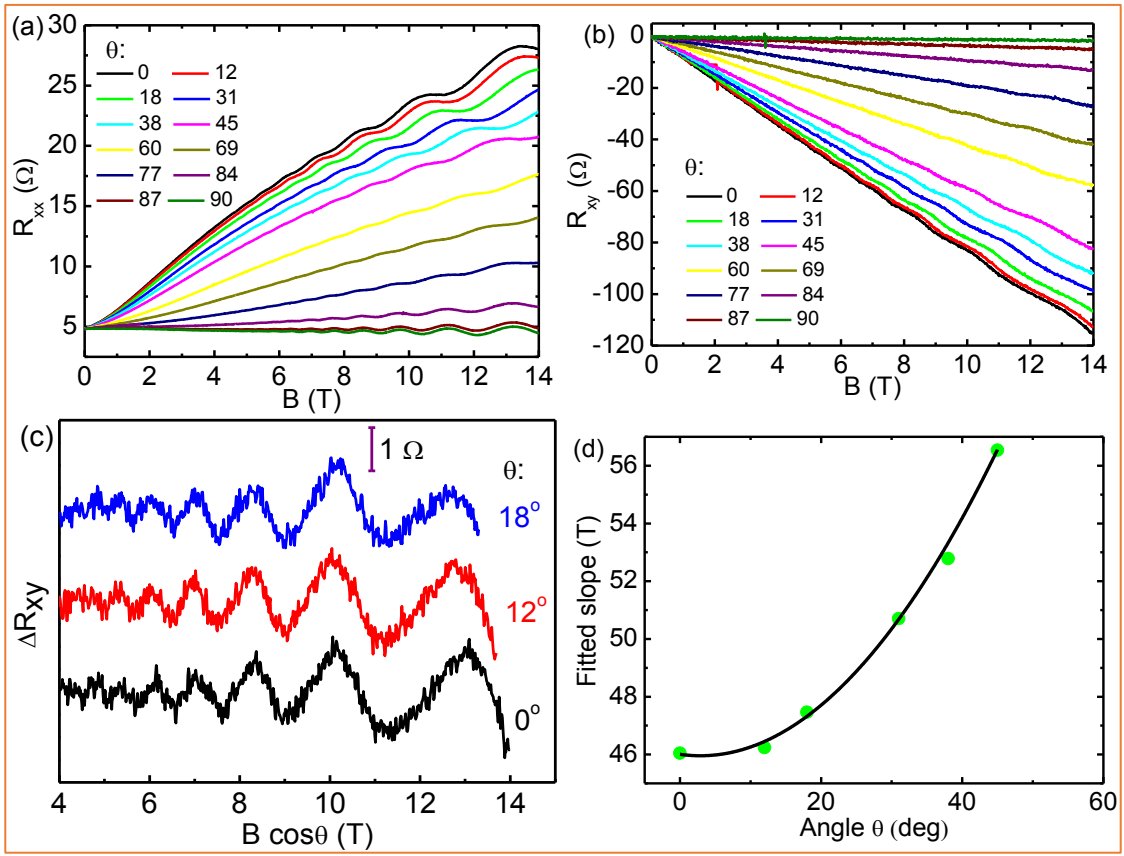

**Figure 4.** (**a**) Longitudinal resistance $R_{xx}$ and (**b**) Hall resistance $R_{xy}$ as a function of the applied magnetic field *B* under different orientations described by the angle θ, at a temperature of 2 K. (**c**) Oscillations of the Hall resistance at 2 K and at different angle θ. (**d**) Fitted slope of the Landau level fan diagram derived from $R_{xy}$ as a

function of the angle θ; the solid line gives a good fitting with $1/cos\theta$.[43]

Angle dependent magneto-transport measurements show that the positive magnetoresistance background, which is the contribution from three-dimensional bulk transport, remarkably decreases as the out-of-plane field component gradually disappear (Fig.4a). To avoid the complexity that top/bottom and sidewalls simultaneously take part in transport, we measured the transversal resistance $R_{xy}$ (2 K), which should not be affected by the sidewall conductance (Fig.4b). Clear oscillation features have been observed (Fig.4c). We derived the Landau level fan diagrams from results in Fig.4b, we found that the derived slopes can be well fitted by $1/cos\theta$ (Fig.4d). Therefore we can confirm that the observed SdH oscillations avoid influences from three-dimensional bulk transport, and the oscillations under in-plane field should originate from the sidewalls. Moreover, such sidewall associated SdH oscillations successfully avoid disturbances from bulk transport, providing an easy method to detect surface states in topological insulators.

## 3 Transport properties of topological Dirac semimetal $Cd_3As_2$

**3.1 Giant negative magnetoresistance induced by the chiral anomaly**

$Cd_3As_2$ is one kind of the so-called three-dimensional Dirac semimetals, which are also described as three-dimensional analogues of graphene. These materials are unusual quantum materials containing massless Dirac fermions[19]. By breaking the time-reversal symmetry or spatial inversion symmetry, the Dirac semimetal is believed to transform into a Weyl semimetal with chiral anomaly[34-39,53]. In single crystal $Cd_3As_2$ nanoplates, we find large negative magnetoresistance induced by chiral anomaly. Moreover, the negative magnetoresistance is closely linked with the carrier density, which is tunable by gate voltage and temperature. Our finding is helpful to understand Weyl fermions in Dirac semimetals.

The $Cd_3As_2$ nanoplates were prepared by chemical vapor deposition method. The exposed surface of the nanoplate samples was (112) plane, identified by TEM results (Fig. 5a). The nanoplates possess a thickness of several hundred nanometers (Fig. 5b). The $Cd_3As_2$ samples were then fabricated to contact with Au electrodes on an oxide Si substrate, serving as the back gate. Four-probe measurements were accepted in the experiments (Fig.5c inset). The temperature dependence of resistivity of the samples demonstrates semiconducting-like behavior, which is different from the metallic behavior in $Cd_3As_2$ bulk crystal (Fig.5c). This is mainly attributed to the deviation of carrier density. The $Cd_3As_2$ samples has low carrier density and its Fermi level is close to Dirac points, where the holes in valence bands can be easily thermally activated to conduction bands in high temperature. So the resistivity of $Cd_3As_2$ increases as the temperature decreases due to the reduced thermal activation. After decreasing to a critical low temperature, the thermal energy is not enough for holes to activate across the gap, leading to the metallic behavior. Unlike previous nanowire, the $Cd_3As_2$ bulks usually maintain high carrier density and the Fermi level is always above Dirac points, where the metallic ρ-T mechanism is dominant. As shown in Fig. 5d, the (MR) of Cd$_3$As$_2$ nanoplates exhibits a notable negative MR with in-plane $B$, even under room temperature. The negative MR also emerges in Cd$_3$As$_2$ nanowire samples with $B \parallel E$.

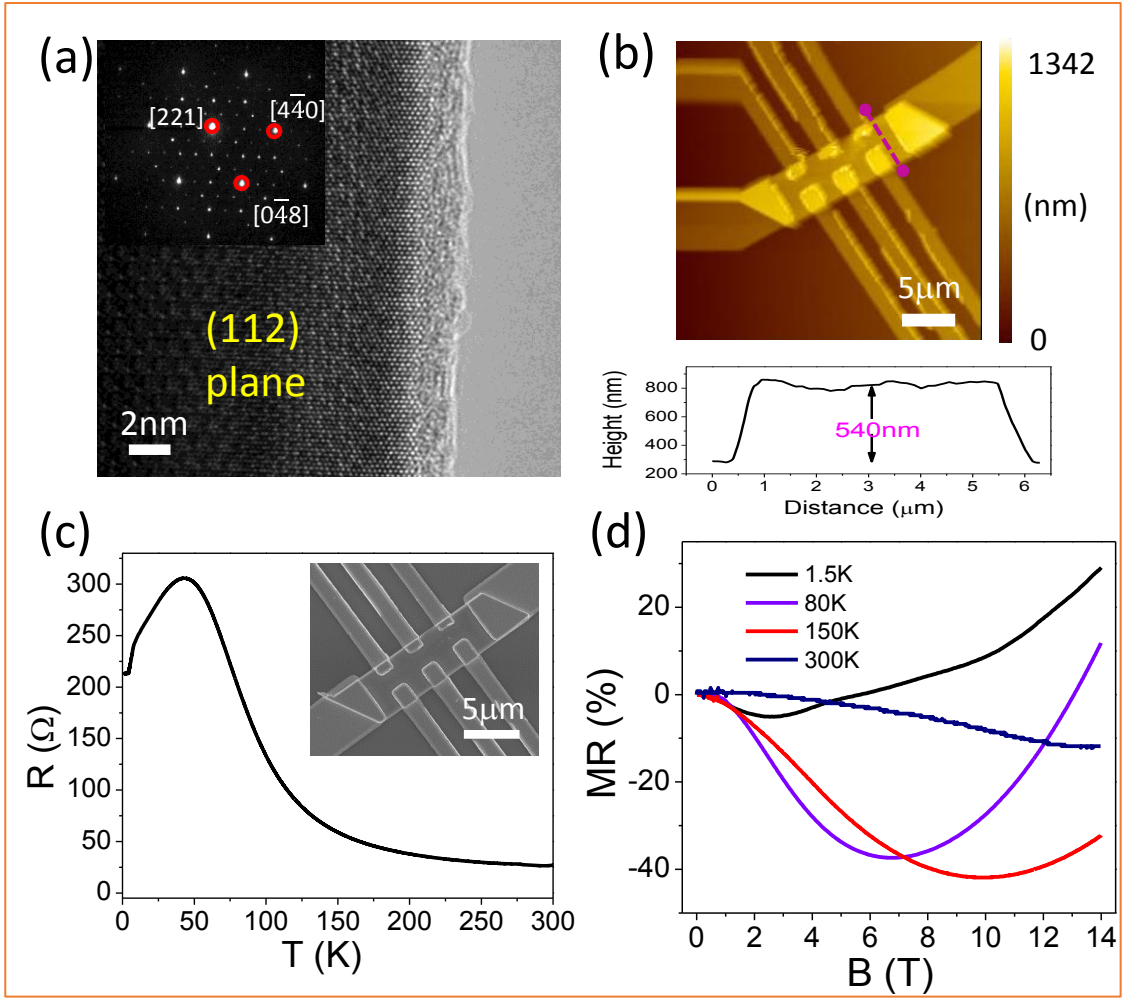

**Figure 5.** (**a**) TEM image of a typical nanoplate that identifies its largest surface to be a (112) plane. Inset: the corresponding diffraction pattern. (**b**) AFM measurement of the nanoplate device with height about 540nm. (**c**) Temperature dependence of resistance of the nanoplate device and its SEM image (inset). (**d**) Plots of MR =100% ×(R(B)/R(0)-1) at variable temperatures. The magnitude of the negative MR reaches maximum ~-41% at ~ 150 K. [40]

The negative MR is mainly induced by the chiral anomaly effect. Cd$_3$As$_2$, as a three-dimensional Dirac semimetal, its Dirac point is composed of two overlapping Weyl nodes with opposite chirality (left handed or right handed) (Fig. 6a), which would be separated in the magnetic field[54,55]. Applying an extra parallel electric field, the two kinds of Weyl fermions would have unequal chemical potential ($\mu_R \neq \mu_L$) (Fig. 6b). In such situation, the continuity equation of the Weyl nodes is

$$\nabla \cdot j^{R,L} + \partial_t \rho^{R,L} = \pm \frac{e^3}{4\pi^2 \hbar^2 c} \boldsymbol{E} \cdot \boldsymbol{B}. \qquad (1)$$

The chemical potential imbalance may induce a charge pumping from one Weyl node to another, leading to a chiral imbalance, and a net current can be generated as

$$j_c = j_c^R - j_c^L = \frac{e^2 B}{4\pi^2 \hbar^2 c}(\mu^R - \mu^L). \quad (2)$$

The chiral current is parallel to the direction of electric field, resulting in the negative MR. By analyzing the negative MR of $Cd_3As_2$, we are capable to verify the existence of chiral characters in $Cd_3As_2$ electronic structure.

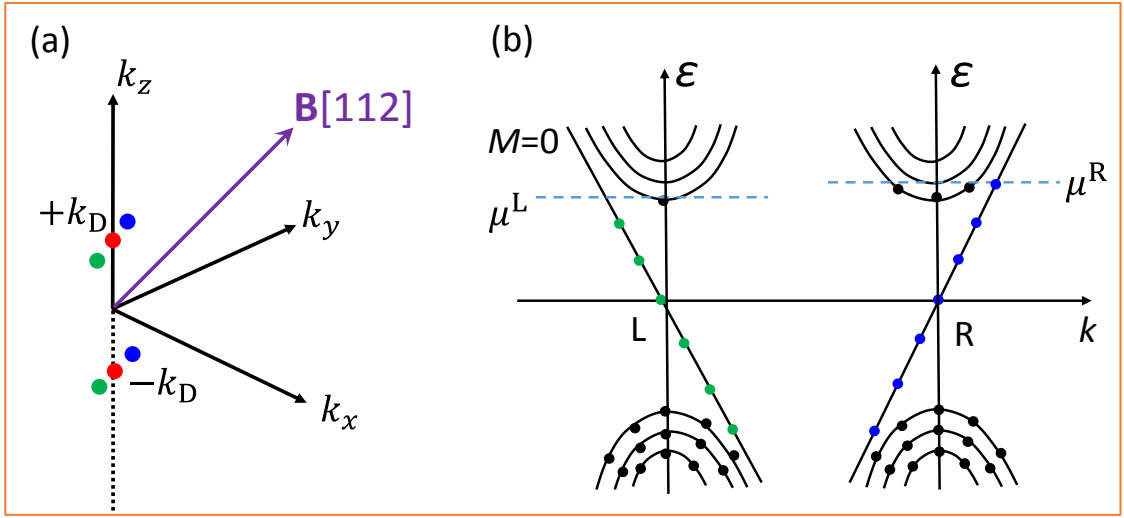

**Figure 6.** (**a**) Illustration of split of Dirac points into Weyl nodes under an external field in [112] direction, $\pm k_D$ represent the Dirac points in the Brillouin zone of $Cd_3As_2$ as marked by red dots, while the green (blue) dots correspond to the split left-handed (right-handed) Weyl nodes along the direction of magnetic field. (**b**)Schematic illustration for the chiral anomaly in Weyl semimetals, L and R correspond to the left-handed and right-handed Weyl nodes respectively. A magnetic field parallel to the electric field generates an imbalance between the two opposite Weyl nodes and thus leads to negative MR, M=0 corresponds to the lowest Landau level.[40]

From the experimental results, we found that the MR curves in low temperatures have two minimum points ($B_{c1}$ and $B_{c2}$)[40]. The first minimum point is ascribed to the fact that the low carrier densities allows the electrons entering into the lowest Landau level at a relatively weak field, leading to an inflection in the MR curve. At low temperatures, the MR increases with magnetic field due to the splitting of conduction bands[53,56]. But with temperatures increasing, the thermal broadening of Landau levels results in the attenuation of upturned MR, leading to the second minimum. Moreover, it is worth to mention that the magnitude of the negative MR and the $B_c$ are closely related to the carrier density. The shift of Fermi level with gate voltage would change the carrier density, resulting in different $B_c$ and magnitude of negative MR. Similarly, with increasing temperature, thermal activation leads to a rise in the carrier density and affects the MR curves. Research suggests that the magnitude of the negative MR decreases with the increase of carrier density, while the critical magnetic field $B_c$ increases corresponding to increasing carrier density.

**3.2 Aharonov-Bohm oscillations in Dirac semimetal $Cd_3As_2$ nanowires**

Two-dimensional Surface states also exist in $Cd_3As_2$ system. It has been predicted that topological surface states with Fermi arcs exist on the surface, which has been supported by angle-resolved photoemission spectroscopy experiments[53,57]. However, the transport evidences revealing the surface states are still in absence, on account of the highly conductive bulk channels. In this section, we show Aharonov–Bohm (A-B) oscillations detected in individual single-crystal $Cd_3As_2$ nanowires, providing transport evidence of the topological surface states in three-dimensional Dirac semimetals.

The single crystal $Cd_3As_2$ nanowires were synthesized by chemical vapor deposition method. The $Cd_3As_2$ nanowires were grown along the [112] direction with single crystalline nature, confirmed by the high-resolution transmission electron microscopy image (Fig.7a). In Fig.7b we display the diagram of a typical device and the measurement configuration.

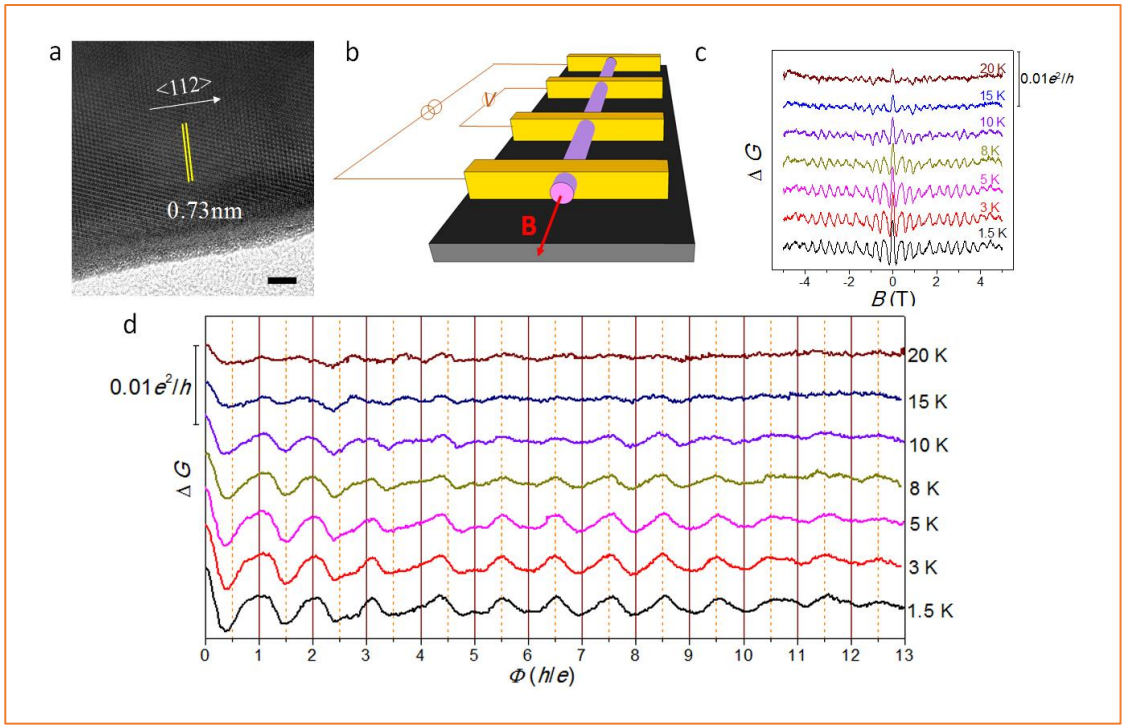

**Figure 7** (**a**) HRTEM image of a typical nanowire indicates <112> growth direction with interplanar space ~ 0.73 nm. Scale bar: 5 nm. (**b**) Schematic diagram of the four-terminal device with applied magnetic field aligned with the length. (**c**) The oscillations in conductance as a function of magnetic field after subtracting background. (**d**) Oscillating conductance as a function of magnetic flux in unit of $h/e$ at variable temperatures.[41]

For nanowires with perimeter comparable to the mean free path of carriers, the quantum confinement enforces the surface energy bands into discrete sub-bands. As sweeping the magnetic field parallel to the nanowire direction, the density of states (DOS) at the Fermi level is periodically altered as sub-bands cross the Fermi level, leading to periodic oscillations of conductance. The oscillating period is described by the $\Phi/\Phi_0$, where $\Phi$ is the magnetic flux threading the nanowire cross section and $\Phi_0=h/e$. The A-B oscillations have been reported in topological insulator nanoribbons and nanowires[58-62].

Applying magnetic field along with [112] direction of $Cd_3As_2$ nanowire, notable oscillations can be detected on a negative magnetoresistance background (Fig. 7c). The oscillation features nearly vanish after heating up to 20 K, indicating its quantum nature. The negative MR is believed to originate from the chiral anomaly. In low field regime ($|\mathbf{B}|<0.3$ T), the MR exhibits an obtuse dip because of weak anti-localization, which is conforming to the existence of strong spin–orbit interaction[53]. The fast-Fourier transform (FFT) on $\Delta G(\mathbf{B})$ extracts the main peak at ~2.55 T$^{-1}$, corresponding to an individual 0.38 T period. If the oscillations originate from the surface states, the period yields a cross-section area of the nanowire ~0.11 μm$^2$. By direct measurements the cross-section area of this nanowire is exactly estimated to be 0.11±0.005 μm$^2$, which identifies the observed quantum oscillations are from the surface state associated A–B effect. We also find that the FFT peak amplitude obeys to $T^{-1}$ law in temperature range 5–20 K, which indicates the ballistic transport nature in circumferential direction[61]. The deviation below 5 K is proposed to be due to the weak influence of thermal broadening on the electron wave packets[63]. The critical temperature for the deviation of $T^{-1}$ dependence is calculated to be ~7.2 K, which is close to the experimental observation (~5 K).

Surprisingly, a phase-shift takes place in the range from 1 to 2 T. The conductance peaks are at the integer quantum flux ($\Phi$=h/e, 2h/e) before the shift, while the conductance peaks are stable at the half-integer quantum flux ($\Phi$=(5+1/2)h/e, (6+1/2)h/e,…) after the phase-shift regime (Fig.7d). The half-integer conductance peak behavior is very similar to that observed in topological insulators, which was ascribed to the $\pi$ Berry phase of the surface states[58-62]. We propose that the degeneracy lifting of the two Dirac nodes under an external magnetic field will provide an additional Berry phase $\pi$ for the surface states, leading to the observed phase-shift. Without magnetic field, the two bulk Dirac nodes (each one contains two degenerate Weyl nodes) are connected by the Fermi arc surface states. The time reversal symmetry is broken after applying an external magnetic field, and the two Dirac points are split into two pairs of Weyl nodes along the magnetic field direction[54]. The Weyl nodes (right-handed chirality and left-handed chirality) are connected by surface Fermi arc respectively after applying magnetic field. The chirality-non-trivial surface states are non-degenerate. An additional Berry phase $\pi$ emerges for electrons after cycling around one Weyl node in the low-energy band, because the degeneracy has been lifted. Moreover, our analysis of Shubnikov-de Haas oscillations also supports the presence of non-trivial $\pi$ Berry phase[41].

The A-B oscillation is actually the periodic variation of DOS at the Fermi level. Therefore, the occupation of sub-bands can influence the behavior of oscillation in aspect of magnitude and phase. By employing a back-gate with 285 nm SiO$_2$, we are able to modulate the sub-band occupation, enabling a deeper understanding of this phenomenon. We found that the gate voltage can effectively modulate oscillation behavior of Cd$_3$As$_2$ nanowires. Moreover, the Lifshitz transition can be achieved by increasing the Fermi wave vector $k_F$. The two bulk Dirac points are separated by $2k_D$ in momentum space. When $k_F$ is larger than $k_D$, the system is back to degenerate, leading to the so-called Lifshitz transition[65]. After the transition, the π Berry phase will disappear, and the peaks of ΔG(**B**) will return to the exact integer numbers. We observed normal A-B oscillations at $V_g = 60V$, with ΔG(**B**) exactly peaking at Φ=5,6,7,8 h/e, just like the theoretical predictions.

For the thick nanowire devices, the carriers along the nanowire perimeter can be in the diffusive regime. Fig.8a illustrates the SEM image of a thick nanowire with diameter ~ 200 nm. Prominent oscillations can be easily discerned, as shown in Fig. 8b. According to the cross-sectional area of the nanowire, it is found that the oscillation periods with *h/2e* flux exist. These anomalous *h/2e* oscillation periods may be attributed to the Altshuler–Aronov–Spivak (AAS) effect in diffusive transport regime. It's worth noting that the oscillation peaks at $\left(n+\frac{1}{2}\right)h/2e$ (*n*=2,3,4), revealing a phase shift for the *h/2e* oscillations which requires further investigations.

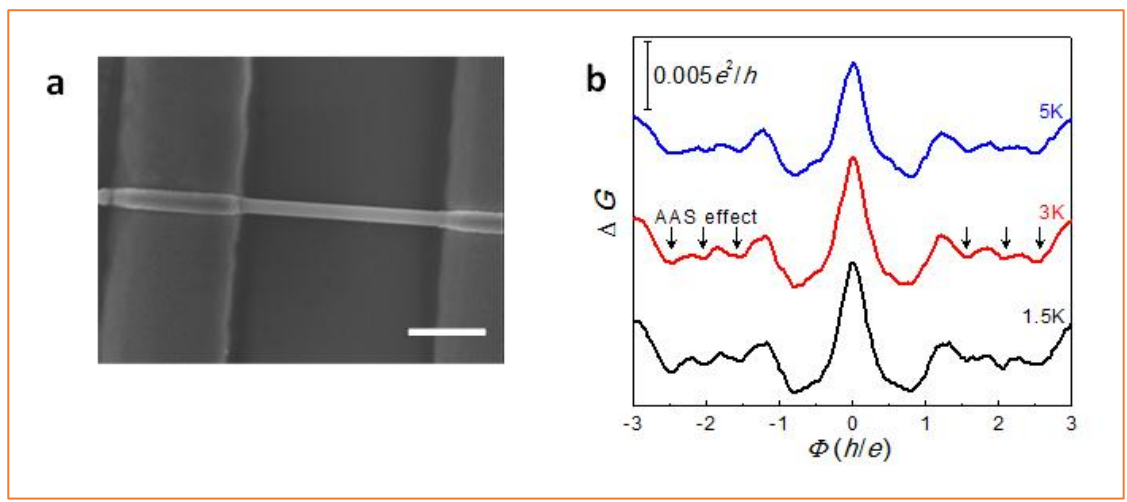

**Figure 8.** (**a**) SEM image of a thick nanowire with diameter about 200 nm. Scale bar: 1 μm. (**b**) The conductance oscillations as a function with magnetic flux, showing the AAS effect. [41]

### 3.3 Two-carrier transport in Cd$_3$As$_2$ nanoplates

In semimetals (graphite and bismuth), large MR under low magnetic fields is always ascribed to the two-carrier transport[66,67]. The saturation of MR under high field is due to the breaking of electron-hole balance[68]. Previous reports have figured out that large MR exists in high carrier concentration Cd$_3$As$_2$ bulk materials[69,70], and Cd$_3$As$_2$ with Fermi level near the Dirac point as well[68]. Therefore, Cd$_3$As$_2$ may be an ideal system to investigate the relationship between the large MR and two-carrier transport. Here we show the temperature and gate voltage-dependent magneto-transport properties of $Cd_3As_2$ nanoplates with Fermi level closed to the Dirac point. Taking advantage of low carrier density in the $Cd_3As_2$ nanoplates, the ambipolar field effect is realized. The Hall anomaly demonstrates the two-carrier transport accompanied by a transition from n-type to p-type conduction with decreasing temperature. The magnetoresistance exhibits a large unsaturated value up to 2000% at high temperatures, which is ascribed to the electron-hole compensation in the system. The results are valuable for understanding the experimental observations related to the two-carrier transport in Dirac semimetals.

High-quality $Cd_3As_2$ nanoplates were synthesized via a facile CVD method. The $Cd_3As_2$ nanoplates are single crystalline and the top surface are (112), consistent with the natural cleavage plane of $Cd_3As_2$ single crystal. To investigate the magneto-transport properties of the as-grown $Cd_3As_2$ nanoplates, back-gate devices with a standard Hall-bar geometry were fabricated, as schematically illustrated in Fig.9a. A typical temperature dependence of resistivity shows semiconducting characteristic (Fig.9b), which is ascribed to the low carrier density in the samples. The activation energy $E_a$ is extracted to be 19.53meV according to the Arrhenius plot at high temperatures (inset in Fig.9b).

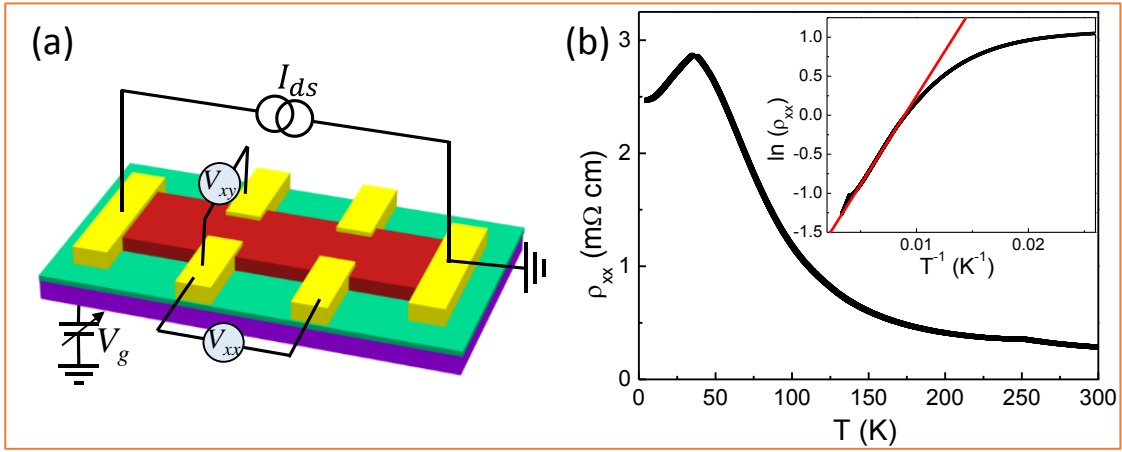

**Figure 9.** (a) Schematic diagram of the $Cd_3As_2$ nanoplate Hall device with a back gate terminal. (b) Temperature dependent longitudinal resistivity $\rho_{xx}$ of $Cd_3As_2$ nanoplate. The activation energy of 19.53 meV is deduced from the Arrhenius plot of $\rho_{xx}$ in the inset.[44]

The evolution of the Hall curves with temperature (Fig.10a) clearly shows a transition from p-type to n-type conduction with increasing temperature. The observed Hall anomaly is a characteristic of two-carrier transport, which can be described by the two-carrier model[72-74]:

$$\rho_{xy} = \frac{1}{e}\frac{(n_h\mu_h^2-n_e\mu_e^2)+\mu_h^2\mu_e^2B^2(n_h-n_e)}{(n_h\mu_h+n_e\mu e)^2+\mu_h^2\mu_e^2B^2(n_h-n_e)^2}B, \quad (3)$$

$$\rho_{xx} = \frac{1}{e}\frac{(n_h\mu_h+n_e\mu_e)+(n_e\mu_e\mu_h^2+n_h\mu_h\mu_e^2)B^2}{(n_h\mu_h+n_e\mu_e)^2+\mu_h^2\mu_e^2B^2(n_h-n_e)^2}, \quad (4)$$

where $n_e$ ($n_h$) and $\mu_e$ ($\mu_h$) are the carrier density and mobility of electrons (holes), respectively. The magnetic field dependence of the Hall resistivity can be well fitted according to eq 3 at representative 80 and 100 K, as shown in Fig.10b. While at 200 and 300 K, the MR shows a quadratic behavior without any tendency to saturation. Largest MR (defined as $\frac{\rho_{xx}(B)-\rho_{xx}(0)}{\rho_{xx}(0)} \times 100\%$) up to 2000% is observed at 200 K (Fig.10c). The observed large unsaturated MR at high temperatures may be due to the thermally activated electrons, leading to the two-carrier transport. As shown in Fig.10d, the MR at different temperatures could be rescaled by the Kohler's plot[72,74]:

$$\frac{\Delta R_{xx}(B)}{R_{xx}(0)} = F\left(\frac{B}{R_{xx}(0)}\right). \quad (5)$$

Moreover, the gate-tunable MR behaviors are consistent with the broken electron-hole compensation induced by the gate voltage. A nearly electron-hole compensation state gives rise to the largest MR at 200 K. shows the electron density increases drastically above 150 K, consistent with the $\rho_{xx} - T$ behavior. In the presence of two parallel conduction paths, the eq 3 can also be used to analyze the Hall data because the two-channel transport can be described in the frame of two-carrier model.

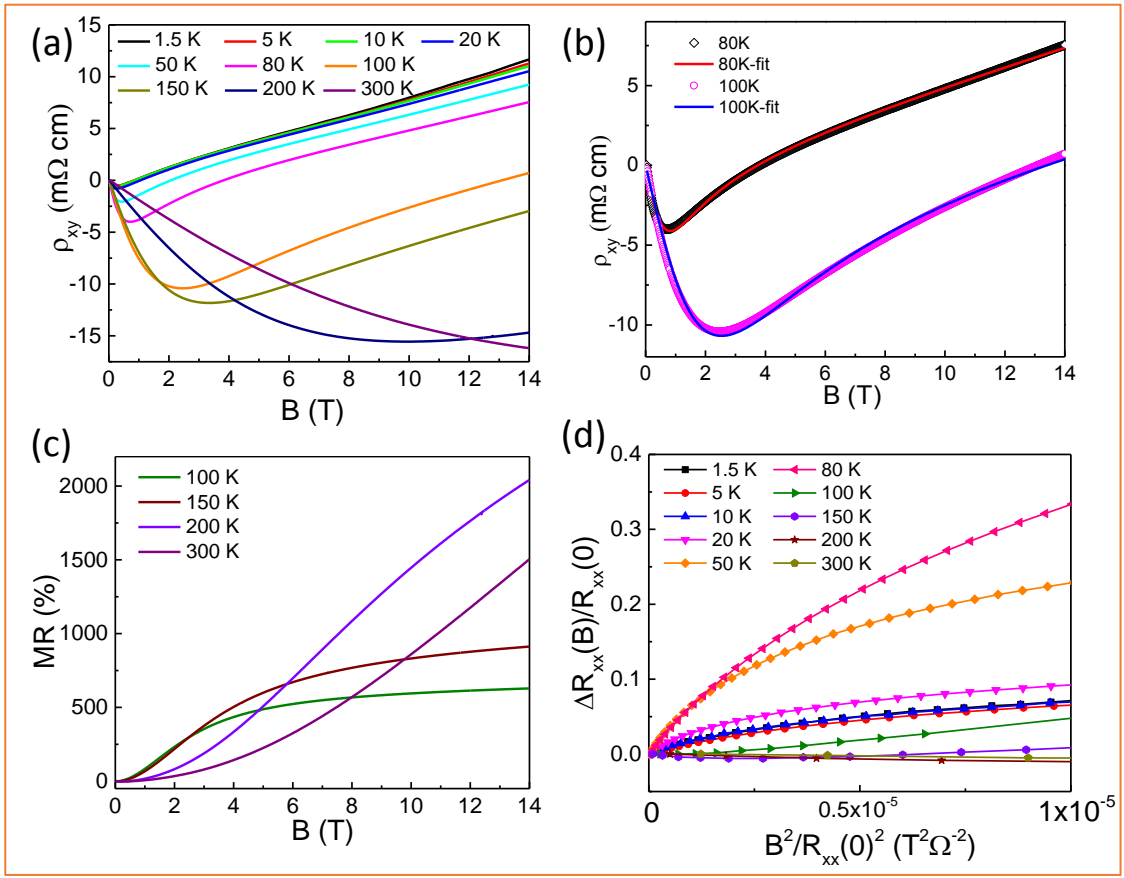

**Figure 10.** (**a**) The Hall resistivity at different temperatures from 1.5 K to 300 K. (**b**) The nonlinear fits of the Hall resistivity based on two-carrier model at representative temperatures of 80 K and 100 K. (**c**) MR measured at different temperatures from 100 K to 300 K. (**d**) The Kohler's plot of the MR curves.[44]

## 4 Gate-tunable tunneling resistance in graphene/TI vertical junctions

Graphene-based Van der Waals heterostructures, formed by stacking graphene with other nanostructures, have become a prosperous avenue of nanoelectronics, which exhibit rich physical phenomena. In this section we review our work on the stacking of two model Dirac materials, graphene and $Bi_2Se_3$. Theoretical predictions have indicated significant modification of graphene electronic structure in such graphene–topological insulator hybrid systems[75-79]. However, experimental investigations of vertical transport properties in such heterostructures remain absent. Here, we show that transport through the vertical heterostructure is dominated by the phonon assisted tunneling process at low temperature, which can be effectively modulated by gate voltage. Quantum oscillations due to the quantized Landau levels in both graphene and $Bi_2Se_3$ surface Dirac fermions are observed, after a magnetic field is applied. Tunneling phenomenon when graphene becomes a quantum Hall insulator is also analyzed. Our discoveries shed light on density of states (DOS) detector based on heterojunctions.

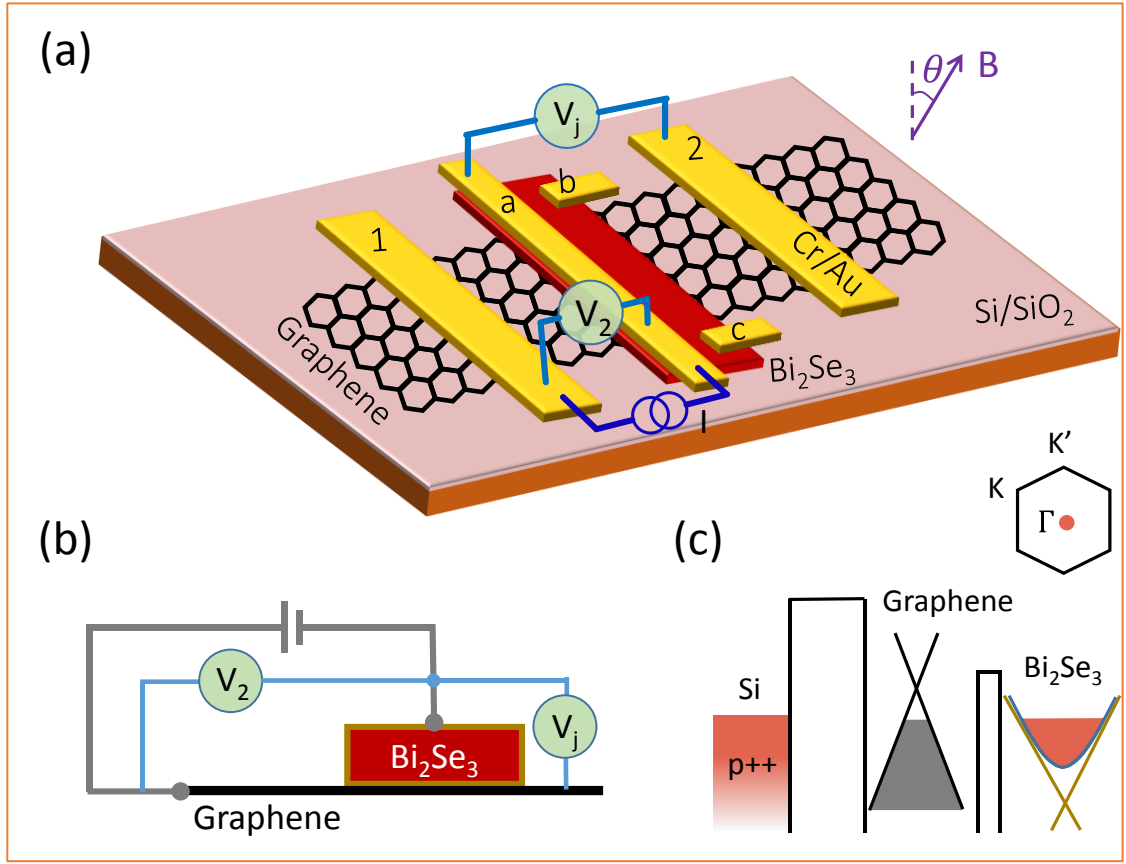

**Figure 11.** (**a**) Schematic of graphene-$Bi_2Se_3$ heterostructure. Graphene is contacted with the electrodes 1 and 2. $Bi_2Se_3$ is contacted *via* the electrodes a, b, and c. The bias current is applied through electrodes 1 and a. (**b**) Side view of the measurement configuration. (**c**) Band diagram of the junction without bias voltage. The upper right panel shows the Brillouin zone of the graphene-$Bi_2Se_3$ hybrid system. The center of the Dirac bands of graphene (or $Bi_2Se_3$) is located at $K, K'$ (or $\Gamma$) points.[45]

To fabricate the hybrid devices, high quality $Bi_2Se_3$ nanoplates were transferred onto monolayer graphene flakes which were prepared by micromechanical cleavage method. Deposition of patterned Cr/Au (50/170 nm) electrodes were followed. Device diagram and measurement configurations are shown in Fig.11a. The resistance $R_j = V_j/I$ is the junction resistance through graphene-$Bi_2Se_3$ heterojunction. The resistance $R_2 = V_2/I$ is a combination of the junction resistance and the underlying graphene resistance. Therefore, the resistance $R_g = R_2 - R_j$ is the in-plane resistance of the graphene. $R_{xy} = V_{xy}/I$ represents the transverse Hall-like resistance of $Bi_2Se_3$. A back gate voltage $V_g$ was applied to tune the Fermi level. The external magnetic field can be tilted by rotating the sample holder.

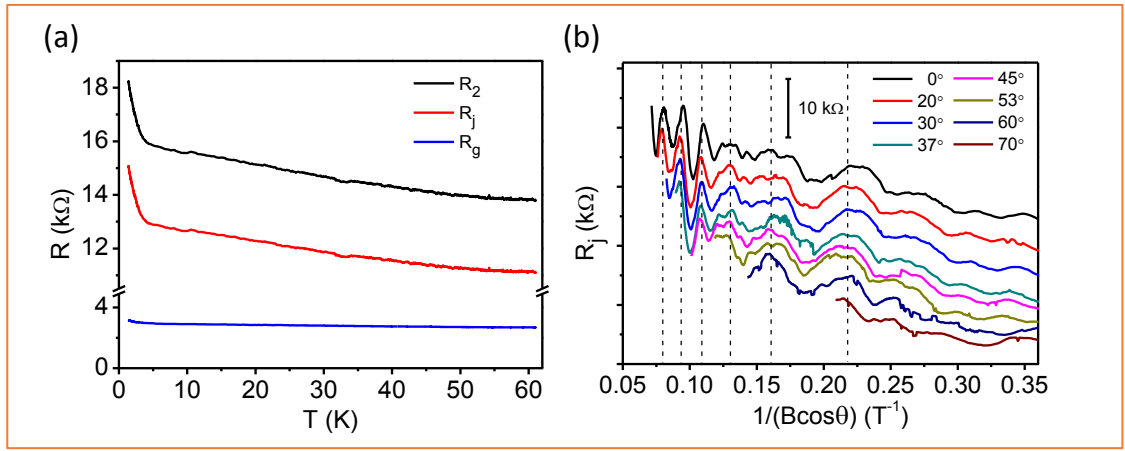

**Figure 12.** (a) $R_2$ (black line), $R_j$ (red line) and their difference $R_g$ (blue line) as a function of temperature. (b) $R_j$ versus $1/B_\perp = 1/(B\cos\theta)$ for different values of $\theta$. Curves are shifted for clarity.[45]

Firstly we analyze the temperature-dependent behavior of the junction resistance. The junction resistance $R_j$ exhibits sensitive termperature dependence, while the graphene resistance $R_g$ is weakly correlated to temperature (Fig.12a). The tunneling efficiency is limited by the graphene/$Bi_2Se_3$ junction potential barrier $\Delta$, thus thermally excited carrier density can be described by $n(T) \sim e^{-\Delta/k_B T}$ for high temperatures, the fitting results also give a $\Delta = \sim 0.35$ meV (i.e., $\sim 4$ K)[45]. At low temperatures (T < 4 K), thermal activation energy cannot excite carriers to exceed the barrier, transport is dominated by tunneling current between graphene and $Bi_2Se_3$. Because of the large momentum mismatch (Fig.11c), tunneling should be assisted by phonons, which provide momentum to inject an electron from $Bi_2Se_3$ to graphene. The tunneling in graphene/BN heterostructures has also been reported to be assisted by phonons in graphene[80].

The tunneling heterostructures also exhibit novel quantum oscillations (Fig.12b). The stable peak positions versus $1/cos\theta B$ indicate the 2D nature of the quantum oscillations. Applying relatively small magnetic fields, the oscillations of junction resistance $R_j$ are mixed with the contributions from both $Bi_2Se_3$ and graphene, displaying complex structures. While under high magnetic field, the oscillations are from the quantized Landau levels of $Bi_2Se_3$ surface states, because the underlying graphene stays in quantum limit when gate voltage is zero. The Dirac point of graphene is at $V_g \sim 5V$, the carrier concentration of the graphene is as low as $3.6 \times 10^{11}\ cm^{-2}$ at zero gate voltage, so in such situation graphene is easily transferred to quantum limit by high enough magnetic field. These results suggest that such a tunneling junction can be used as a DOS detector.

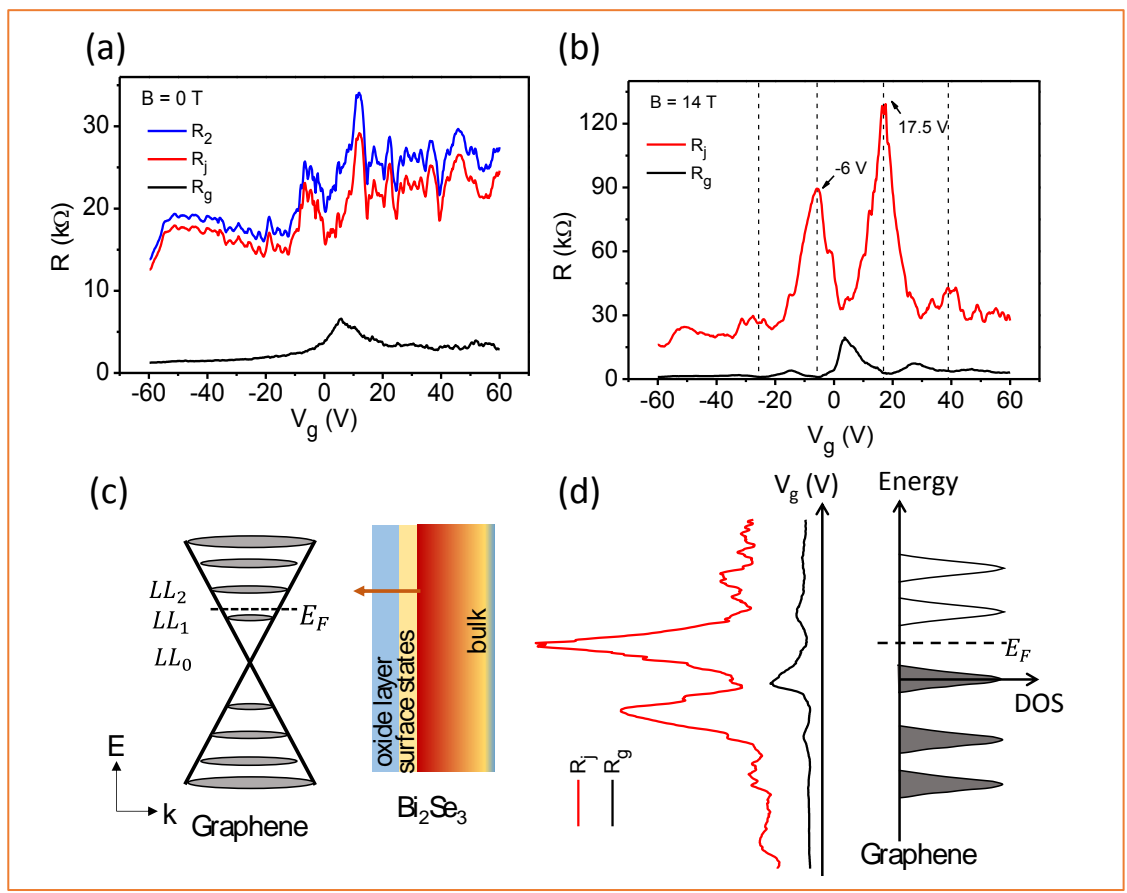

**Figure 13.** (**a**) Resistance *versus* gate voltage without magnetic field. (**b**) Transfer curves of $R_j$ and $R_g$ under $B$ = 14 T. The black dashed lines are guided for eyes to show the alignment. (**c**) Schematic band diagram of the tunneling process under magnetic field. (**d**) The quantum oscillations under $B$ = 14 T are presented to compare with the DOS structure of graphene under magnetic field of 14 T.[45]

The back gate modulation offers a direct method to investigate the tunneling behavior. Under zero magnetic field, the transfer curve of $R_g$, the calculated resistance ($R_g = R_2 - R_j$) of the underlying graphene (Fig.13a), shows typical field-effect properties of a graphene transistor with a clear Dirac point ($V_g \sim 5V$). Besides, the Fermi level of $Bi_2Se_3$ is not sensitive to the gate voltage because of the screening effect from the underlying graphene flake. When the graphene becomes a quantum Hall insulator, exotic phenomena are observed from the graphene–$Bi_2Se_3$ hetero-junctions. Under a large perpendicular field of 14 T, the $R_j$ peaks are exactly located at the positions with the graphene resistance $R_g$ valleys (Fig.13b). That is saying, the tunneling resistance maximum and the graphene resistance minimum always come simultaneously. Considering these results, the gate tunable tunneling resistance $R_j$ is the consequence of the changing Fermi level in graphene and the nearly fixed Fermi level in $Bi_2Se_3$. As the Fermi level in graphene is tuned far away from its Dirac point, the available DOS for tunneling increases greatly, resulting in a decreasing $R_j$ (Fig.13c). After transforming into a quantum Hall insulator, the quantum Hall states in graphene provide large DOS at unfilled Landau levels (LLs) for tunneling carriers. However, there are almost no available tunneling states within the gaps between LLs. Therefore, the junction resistance $R_j$ valleys locate at the graphene LLs, while the peaks locate at the gaps where LLs are fully occupied (Fig.13d).

## 5 Recent progress on 2D topological transport

### 5.1 Topological edge transport

Nowadays, edge state transport is particularly important in low-consumption electronic devices due to its non-dissipative transport property. For quantum materials with non-zero Chern number, which is always induced by time-reversal symmetry breaking, topological nontrivial surface/edge states exist on the boundary of the materials, where backscattering is entirely suppressed in these states, leading to realization of non-dissipative transport. On the other hand, various kinds of Hall effects are critical phenomena for existence of these nontrivial states in 2D systems, as a result, QHE[2,7-16], quantum spin Hall effect (QSHE)[81-84] and quantum anomalous Hall effect (QAHE)[25-27] are focus of experimental investigations. These novel phases always show a bulk insulating behavior together with perfectly conducting edge transport. The QHE, a quantized version of the Hall effect, was observed in various 2D electron gas systems over 30 years. For recent years, experimental realizations of QSHE and QAHE become the fantasy pursued by researchers. Here, we make a summary about recent theoretical and experimental investigations on 2D topological edge transport.

### 5.1.1 Quantum spin Hall effect

The QSHE, a quantized response of a transverse spin current to an electric field, hosts two counter-circulating states with opposite spin projection at each edge. The intrinsic SOC is proposed to give rise to the QSH effect. Protected by time reversal symmetry, QSH state has a topological stability that is insensitive to weak disorder and impurities. QSH effect was firstly predicted to occur in HgTe quantum wells by Bernevig, B.A *et al.* and then was observed experimentally by Roth A *et al*[81,84], starting a new epoch of QSH. Theoretically, QSH is also predicted in graphene system with strong SOC[83]. However, the intrinsic SOC in graphene is shown to be too weak to realize the QSH effect under present experimental conditions, in which case enhancement of SOC is instantly required. We noticed the proximity effect is possible to enhance the SOC in graphene system, Inspired by works based on graphene Van der Waals heterostructures and hydrogenated graphene[85,86], the proximity effect is possible to enhance the SOC in graphene. Today, QSH still has a broad prospect for development in spintronics, owing to the fact that spin current in QSH is carried entirely by the helical edge states without dissipation.

**5.1.2 Quantum anomalous Hall effect**

Unlike the conventional QHE, which is induced by a strong magnetic field, the QAHE describes the non-dissipative quantized Hall transport in the absence of external magnetic fields. The QAH is a possibility in strong spin-orbit coupled magnetic 2D electron system. The time-reversal symmetry is broken by simultaneous magnetism. The QAHE is predicted to exist in magnetized TI systems[25], and in 2013, Chang, C.Z *et al.* has firstly observed the QAH experimentally in chromium-doped topological insulator $(Bi,Sb)_2Te_3$[26]. And then in 2015, robust quantum anomalous Hall state was realized in V-doped $(Bi,Sb)_2Te_3$[27].Recently, it is theoretically predicted that QAH can be realized in graphene coupled to an antiferromagnetic insulator by introducing Rashba SOC and an exchange field[87,88]. The fabrication of materials showing the QAH phase often has enormous potential since they are expected to show a perfect edge conductance as the quantum Hall state, but without the drawback of having to apply large magnetic fields.

**5.2 Topological valley transport**

Topological materials with time-reversal symmetry breaking may possess novel phenomena like QHE or QAHE, revealing non-dissipative edge transport. In the contrast, if the inversion symmetry is broken, transport related to valley pseudospin may emerge, leading to the so-called "valleytronics"[2,89]. The valley current also maintains long-range character, indicating a potential vista for next-generation information devices. In this section we review the progress on 2D topological valley transport, including valley Hall effect (VHE) observed in $MoS_2$, and valley current transport in single layer and bilayer graphene.

**5.2.1 The valley Hall effect**

A system containing non-zero Berry curvature may obtain an additional velocity term: $v_n(\mathbf{k}) = \frac{\partial \epsilon_n(\mathbf{k})}{\hbar \partial \mathbf{k}} - \frac{e}{\hbar} \mathbf{E} \times \mathbf{\Omega}_n(\mathbf{k})$. In additional to the typical band dispersion contribution, an anomalous term due to the Berry curvature emerges, the extra term is always transverse to the electric field, leading to a Hall current[2]. For the system with inversion symmetry breaking, the Berry curvature in different valleys have opposite signs under the protection of time-reversal symmetry, thus non-zero valley conductivity can exist in the transverse direction, net valley degree accumulation is expected on both transverse side. This phenomenon is called the "valley Hall effect". To experimentally realize the VHE, suppressed valley relaxation must be achieved. The spontaneous inversion symmetry breaking and strong spin-orbit coupling make $MoS_2$ an ideal platform to realize VHE[90]. K.F. Mak *et al.* firstly reported the experimentally observation of VHE in monolayer $MoS_2$[91]. Circularly polarized light was used to create chemical potential imbalance in the two valleys of $MoS_2$, inverting valley signal to an observable electric signal, anomalous Hall signal induced by VHE was substantially detected.

**5.2.2 The valley transport in graphene**

Theoretical analysis has shown that the graphene with inversion symmetry breaking can also exhibit the VHE[92]. However, pristine graphene (single layer or bilayer) is under inversion symmetry protection. The experimental investigations on the valley transport in graphene system have been achieved recently, with the help of improvement on device fabrication.

In 2014, R.V. Gorbachev *et al.* reported large non-local signal detected in single layer graphene with underlying BN substrate[93]. The inversion symmetry of graphene is believed to be broken, as the consequence of A/B sublattice global asymmetry induced by substrate potential. The inversion symmetry breaking leads to non-zero Berry curvature distribution in the Brillouin zone. When the Fermi level is tuned to the energy level where non-zero Berry curvature exists, large non-local signal can be detected from the transport measurements. The anomalously obvious non-local signal is due to the valley Hall effect and the inverse VHE, the latter is the phenomenon where pure valley current gives rise to charge current.

The novel work from R.V. Gorbachev *et al.* not only firstly reported the experimental realization of graphene valley transport, but also provided an efficient method to detect the VHE. In 2015, the valley transport in bilayer graphene is discovered by M. Sui. *et al.* and Y. Shimazaki *et al.* The two groups both utilized perpendicular electric field tuned by a gate voltage to break the inversion symmetry of graphene, strong non-local signal was detected, which was attributed to the VHE and the inverse VHE[94,95]. Their investigations show that electrically driven inversion symmetry breaking is achievable, shedding light on future valleytronics. Furthermore, the AB/BA stacking domain wall in bilayer graphene has also been reported to support a non-diffusional valley current[96]. The bilayer graphene system with broken inversion symmetry may possess a non-zero "valley" Chern number, leading to topological valley edge state existing along the domain wall.

**5.3 Topological Weyl semimetals**

Based on the quantum field theory, the fermions of our universe are divided into three kinds: the Dirac fermions, the Weyl fermions and the Majorana fermions[97]. Interestingly, the condensed-matter systems host quasiparticles possessing the characters of these high-energy physics concepts. The most typical example of Dirac semimetals is graphene, which has a Dirac point protected by time-reversal symmetry and inversion symmetry[10-16]. Extending the 2D linear energy-momentum dispersion relation to 3D momentum space, the 3D Dirac semimetals were recently discovered. Moreover, the Dirac point in 3D Dirac semimetals is composed of two Weyl nodes. The topological Weyl semimetals are the systems where low-energy excitations obey the Weyl equations[57,98-100]. Compared to the Dirac points (four-fold degeneracy), the Weyl nodes are only two-fold degenerate. Thus the Weyl node can only exist in systems with spontaneous symmetry-breaking, because in the systems possessing both the time-reversal and inversion symmetry the band crossings are always four-fold degenerate.

The topology of the Weyl nodes is also defined by the Berry curvature[98-100], an analogy of the magnetic flux in momentum space. Generally speaking, the Weyl nodes can be considered as the "magnetic monopoles" in momentum space, which are the origins or ends of the Berry curvature. The chirality (+1 or -1) of a Weyl node is similar to the "magnetic charge", reflecting the Berry curvature distribution embracing the Weyl node. The Weyl nodes always come in pairs because the Chern number of the Weyl semimetal must be zero, otherwise it becomes a topological insulator[2].

There are a large number of unusual physical features associated with Weyl semimetal, including open Fermi arcs in the surface and various exotic transport phenomena induced by the chiral anomaly. The open Fermi arcs connect the projections of the bulk Weyl nodes with opposite chirality onto the surface, and the Fermi arcs can be understood as the surface states of Weyl semimetals[98]. Under the presence of external electric and magnetic field, the chiral charge of one Weyl node is not conserved, which is the so-called chiral anomaly. Charge pumping effect between Weyl nodes with opposite chirality gives rise to large negative magnetoresistance[40,54-57,97,98].

It is worth noting that the Weyl semimetals discussed above have zero density of state at the zero energy, and they obey the Lorentz invariance in quantum field theory. This kind of Weyl semimetals have been predicted to exist in several symmetry-breaking materials, including TaAs, NbP and other systems[99,100]. By ARPES technique, the existence of Weyl fermions in TaAs, NbAs has been experimentally confirmed[101-105], and chiral-anomaly-induced negative magnetoresistance is also detected in TaAs recently[106].

Recently, a new type of Weyl semimetal has been predicted in $WTe_2$ system[107]. In this type of Weyl semimetal, non-zero density of state at zero energy exists, and such system seems to break through the limit of Lorentz invariance, which is referred as a type-2 Weyl semimetal. The remarkable distinction between the Fermi surfaces of the two types of Weyl semimetals leads to notable differences in the thermodynamics and magneto-transport. The type-2 Weyl semimetal exhibits the chiral anomaly effect with strong anisotropy.

## 6  Conclusion and Perspectives

Topological nontrivial systems provide splendid platforms for uncovering novel physics in mesoscopic scale, and also for practical applications. As a powerful method of physical characterization, transport measurements make significant contributions in exploring these miraculous systems. This topical review summarizes our experimental investigations on transport behaviors in topological insulator ($Bi_2Se_3$), topological Dirac semimetal ($Cd_3As_2$), and TI-graphene heterojunction. We demonstrate fantastic quantum effects of these systems, exhibiting their great potential in realizing next-generation electronic and spintronic devices. We also review significant breakthroughs from other groups, to depict a general status of current research on topological transport. Although great triumphs have been achieved, the future investigations are still in eager expectation.

The past ten years have witnessed great achievements on graphene system. Moreover, this intriguing 2D material still possesses unpredictable academic and application prospect. Researchers begin to enlarge the functionalization of graphene using graphene-based heterostructures by proximity effect. Other than the inversion symmetry breaking and enhancement of spin-orbit coupling discussed above, large exchange field has been introduced into graphene by a magnetic insulator EuS[108]. We believe experimental realizations of QSHE and QAHE in graphene-based heterostructures will come true in the near future. The physical contents and application possibilities are obviously diversified by investigations on the graphene-based heterostructures.

The proximity effect also shows extraordinary talents in other two-dimensional systems. The superconducting proximity effect is believed to induce Majorana fermions in a topological nontrivial system[109], and transform the hybrid system into a "topological superconductor". The controllable realization of such fantastic fermions will start the revolution of quantum computing. These features lead the superconducting proximity effect to become a research focus in recent years. Researchers have successfully detected signature of Majorana fermions in topological superconductor nanowires[110,111], and we notice a recent report has provided evidences of the Majorana zero mode in $Bi_2Te_3$/$NbSe_2$ topological superconductor[112]. The superconductivity in Dirac semimetal $Cd_3As_2$ was also achieved by a point contact[113]. The topological superconductor systems will remain intriguing for a long time, and there is still inestimable amount of physical contents for us to discover.

Many other novel 2D materials are waiting to be revealed. The black phosphorus, which has both band gap and high mobility, is discovered in recent years[114-117]. Researchers have inquired into its application prospect[114,115], and the QHE has been realized in black phosphorus[116,117]. Other new systems, like type-2 Weyl semimetal, are explored continuously. The variety of 2D materials affords us incessant inspirations, and the combination of 2D materials and other systems (for example, 2D material with superconductors or ferromagnets) will actually maximize such variety.

On the other hand, the improvement of device fabrication techniques provides substantial foundation for intensive works. The MBE method has been developed to grow high-quality 2D materials and their Van der Waals heterostructures. The interference of impurity potential can be greatly suppressed by the device encapsulation technology (for example, BN encapsulation)[118], leading to ultraclean device environment. By the help of technological advance, more and more physical phenomena have been discovered experimentally.

We hope the readers of *Chinese Physics B* may acquire inspirations from our review, and our investigations over these years may shed light on future explorations.

**Acknowledgment**

We would like to thank our collaborators sincerely for contributing to the related works presented in this review. This work was supported by MOST (Nos. 2016YFA0300802, 2013CB934600) and NSFC (Nos. 11274014, 11234001).